\documentclass{eptcs}
\providecommand{\event}{IMPEX 2017} % Name of the event you are submitting to
\usepackage{underscore}           % Only needed if you use pdflatex.
\usepackage{graphicx}
\usepackage{listings}
\usepackage {bsymb,b2latex}

\title{Formal Modelling of Ontologies : \\An Event-B based Approach Using the Rodin Platform\footnote{The work reported in this paper has been supported by the ANR project IMPEX ref : Projet-ANR-13-INSE-0001}}
\author{Yamine AIT AMEUR
\institute{IRIT/INPT-ENSEEIHT\\ Toulouse, France}
\email{yamine@enseeiht.fr}
\and
Idir AIT SADOUNE 
\institute{LRI/CentraleSupelec\\Gif Sur Yvette, France}
\email{\quad idir.aitsadoune@centralesupelec.fr }
\and Kahina  HACID 
\institute{IRIT/INPT-ENSEEIHT\\ Toulouse, France}
\email{kahina.hacid@enseeiht.fr}
\and Linda MOHAND OUSSAID
\institute{LRI/CentraleSupelec\\Gif Sur Yvette, France}
\email{\quad linda.mohandoussaid@centralesupelec.fr }
}
\def\titlerunning{Formal Modelling of Ontologies within Event-B}
\def\authorrunning{Yamine Ait Ameur, Idir Ait Sadoune, Kahina Hacid and Linda Mohand Oussaid. }
\begin{document}
\maketitle

%%%%%%%%%%%%%%%%% 
%%%%%%%%%%%%%%%%%
\begin{abstract}
This paper reports on the results of the French ANR IMPEX research project dealing with making explicit
domain knowledge in design models. Ontologies are formalised as theories with sets, axioms,
theorems and reasoning rules. They are integrated to design models through an annotation mechanism.
Event-B has been chosen as the ground formal modelling technique for all our developments.
In this paper, we particularly describe how ontologies are formalised as Event-B theories.
\end{abstract}

%%%%%%%%%%%%%%%%%
%%%%%%%%%%%%%%%%%
\section{Introduction}

Nowadays, it is well accepted that formal ontologies are commonly used as support for the axiomatisation of the knowledge describing a domain of interest. In particular, for domains in the engineering area where concepts are well mastered by the different stakeholders, ontologies play a major role for knowledge exchange and heterogeneity reduction.

Meanwhile, we observe that defining a formal framework for integrating both ontologies represented by knowledge models and design models of particular systems did not draw the attention of many researchers in system engineering.

Approaches like those of \cite{Isola2014}\cite{SCP2015}\cite{tase15}\cite{Isola2016}\cite{mediMohandOussaidA17}\cite{ZayasMA10} supporting the integration of both ontologies and design models contribute to strengthen these design models by offering the capability to design models to borrow knowledge from ontologies, using a particular annotation relationship. As a consequence, the design models are enriched and strengthened with axioms, theorems or invariants issued from the used ontologies.

This paper presents a summary of the work achieved in the context of the French ANR IMPEX research project. Ontologies are formalised as theories with axioms, theorems and reasoning rules. Event-B \cite{Abrial10} has been chosen as the ground formal modelling technique for all our developments.

%%%%%%%%%%%%%%%%%
%%%%%%%%%%%%%%%%%
\section{Event-B formal developments}

The Event-B method \cite{Abrial10}  is a formal method based on first order logic and set theory. It relies on the notions of pre-conditions and post-conditions, weakest pre-condition and the calculus of substitution.  An Event-B model is characterised  by a set of variables, defined in the VARIABLES clause that evolve thanks to events defined in the EVENTS clause. It encodes a state transition system where the variables  represent the state and the events represent the transitions from one state to another.

%%%%%%%%%%%%%%%%%
%%%%%%%%%%%%%%%%%
\subsection{Event-B model}
\begin{figure}[!h]
\begin{center}
\includegraphics[width=11cm]{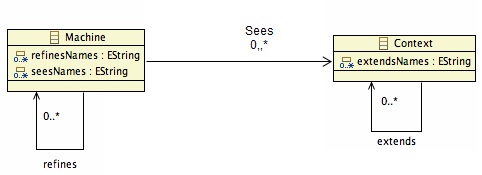}
\caption{MACHINE and CONTEXT relationships}
\label{fig:relation}
\end{center}
\end{figure}

An Event-B model is made of several components of two kinds : Machines and Contexts.  Machines contain the dynamic parts (states and transitions) of a model whereas  Contexts contain the static parts (axiomatisation and theories) of a model. A Machine may be refined  by another one, and a Context may be extended by another one. Moreover, a Machine  sees one or  several Contexts (figure \ref{fig:relation}).

\begin{figure}[!h]
\begin{center}
\scriptsize{
\begin{tabular}{|p{5.8cm}| p{5.8cm}|}
\hline
%\vspace{-0.4cm}
\begin{description}
\CONTEXT{$context\_identifier_1$}
\EXTENDS{$context\_identifier_2$}
\SETS
	\begin{description}
	\ItemX{s }
	\end{description}
\CONSTANTS
	\begin{description}
	\ItemX{ c}
	\end{description}
\AXIOMS
	\begin{description}
	\nItemX{axm}{A(s,c)}
	\end{description}
\THEOREMS
	\begin{description}
	\nItemX{thm}{T(s,c)}
	\end{description}
\END
\end{description}
%\vspace{-0.3cm}
&
%\vspace{-0.4cm}
\begin{description}
\MACHINE{$machine\_identifier_1$}
\REFINES{$machine\_identifier_2$}
\SEES{}
	\begin{description}
	\ItemX{ context\_identifier_1 }
	\end{description}
\VARIABLES
	\begin{description}
	\ItemX{ v }
	\end{description}
\INVARIANTS
	\begin{description}
	\nItemX{inv}{I(s,c,v)}
	\end{description}
\THEOREMS
	\begin{description}
	\nItemX{thm}{T(s,c,v)}
	\end{description}
\VARIANT	
	\begin{description}
	\ItemX{ V(s,c,v)}
	\end{description}
\EVENTS
	\begin{description}
	\ItemX{ <event\_list> }
	\end{description}
\END
\end{description}
%\vspace{-0.5cm}
\\
\hline
\end{tabular}}
\caption{The structure of an Event-B development}
\label{fig:modele}
\end{center}
\end{figure}

A Context is defined by a set of clauses (figure~\ref{fig:modele}) as follows.
\begin{itemize}
\item SETS describes a set of abstract and enumerated types.
\item CONSTANTS represents the constants used by a model.
\item AXIOMS describes, in first order logic expressions, the properties of the attributes defined in the CONSTANTS clause. Types and constraints are described in this clause as well.
\item THEOREMS are logical expressions that can be deduced from the axioms.
\end{itemize}

Similarly to Contexts, a Machine is defined by a set of clauses (figure~\ref{fig:modele}). Briefly, the clauses mean.

\begin{itemize}
    \item  VARIABLES represent the state variables of the model of the specification. Refinement may introduce new variables in order to enrich the described system.
    \item  INVARIANTS describe, by first order logic expressions, the properties of the variables defined in the  VARIABLES clause. Typing information, functional and safety properties are usually described in this clause. These properties shall remain true in the whole model. Invariants need to be preserved by events (by induction). It also expresses the gluing invariant required by each refinement for property preservation.
    \item THEOREMS defines a set of logical expressions that can be deduced from the invariants. They do not need to be proved for each event like for the invariant.
   \item VARIANT introduces a decreasing natural number to ensure termination of  \textit{"convergent"} events.
   \item  EVENTS defines all the events (transitions) that occur in a given model. Each event is characterized by its guard and by the actions performed when the guard is true. Each Machine must contain an "\textit{Initialisation}" event.  The events occurring in an Event-B model affect the state described in VARIABLES clause.
   \end{itemize}
 
 %%%%%%%%%%%%%%%%%
%%%%%%%%%%%%%%%%%
\subsection{Proof obligation rules}   
   Proof obligations (PO) are associated to any Event-B model. They are automatically generated. \textit{The proof obligation generator plugin} in the Rodin platform~\cite{rodin} is in charge of generating them. These PO need to be proved in order to ensure the correctness of developments and refinements. The obtained PO can be proved automatically or interactively by \textit{the prover plugin} in the Rodin platform.  The rules for generating proof obligations to prove the correctness of an Event-B development are given in \cite{Abrial10}. 
   
%%%%%%%%%%%%%%%%%
%%%%%%%%%%%%%%%%%
\section{Need to embed ontologies in formal developments}   
\label{methods}

When design models are produced, designers use domain knowledge in order to formalise the concepts and components of the system to be designed. Usually, this knowledge is not made explicit and is used in an empirical manner. There is no complete formalisation for the reasoning that can be associated to this knowledge

Embedding ontologies in design models in a modular way makes it possible to use ontology concepts and associated reasoning rules in the design models. The interest is to strengthen the models as shown in our previous work \cite{Isola2014}\cite{SCP2015}\cite{tase15}\cite{Isola2016}\cite{mediMohandOussaidA17}\cite{ZayasMA10}. The Event-B method  \cite{Abrial10} has been set up to show how our approach works.

When integrating ontologies and design models, the main difficulty consists in defining a sound integration operation in order to overcome the difficulties resulting from possible semantic gaps that may occur due to the use of ontologies in formal development models. For example, we have adopted the closed world assumption that fits with the studied systems.

To illustrate the approaches proposed in the context of the IMPEX project, we use an extract of the ontology of diplomas described  using OWL formalism \cite{owl2-overview} (figure \ref{fig:modele0}). It defines classes for diplomas ($Diplom$). Other classes subsumed by the diploma class are defined: $Bachelor$, $Master$, $Engineer$ and $Phd.$ This Ontology states that $Master$ and $Engineer$ diplomas are equivalent diplomas, and the concept $Diplomas\_For\_Phd$ is defined as the union of the students that hold an engineer or a master diploma.

\begin{figure}[!h]
\begin{center}
\scriptsize{
\begin{tabular}{|p{12cm}|}
\hline
\lstset{language=XML}
\begin{lstlisting}
<Ontology>
	...
	<Class ID="Diplom"/>
	
	<Class ID="Bachelor">
		<subClassOf resource="Diplom"/>
	<Class> 
	
	<Class ID="Master">
		<subClassOf resource="Diplom"/>
	</Class>
	
	<Class ID="Engineer">
		<subClassOf resource="Diplom"/>
		<equivalentClass resource="Master"/>
	</Class>
	
	<Class ID="Phd">
		<subClassOf resource="Diplom"/>
	</Class>
	
	<Class ID="Diplomas_For_Phd">
		<unionOf parseType="Collection">
			<Class about="Master"/>
			<Class about="Engineer"/>
		</unionOf>
	</Class>
	...
</Ontology>
\end{lstlisting}
\\
\hline
\end{tabular}}
\caption{Extract of the diplomas Owl ontology}
\label{fig:modele0}
\end{center}
\end{figure}

%%%%%%%%%%%%%%%%%
%%%%%%%%%%%%%%%%%
\section{Ontologies as theories}
\label{theories}

As mentioned above, ontologies are formalised as theories integrated to formal system modelling languages. In the context of the IMPEX project, we have identified two approaches to define ontologies
as formal theories. These two approaches use two different modelling processes: shallow \cite{mediMohandOussaidA17} and deep modelling \cite{mediHacidA16,Isola2016}.

%%%%%%%%%%%%%%%%%%%%%%%%%%%%%
%%%%%%%%%%%%%%%%%%%%%%%%%%%%%
\subsection{Shallow modelling: Ontologies as contexts}

The approach that uses shallow modelling consists in modelling the ontology concepts directly in the target modelling language without keeping trace of the structure of the ontology modelling language concepts \cite{mediMohandOussaidA17}. One way to integrate the ontology concepts into a specific formal method development process is to express the ontologies languages constructs into the target formal language by means of transformation rules. In our case, a shallow modelling approach consists in encoding the ontology concepts (classes, properties, ... ) directly in an Event-B context by using abstract sets, constants and axioms.

For example, each class is implicitly a subclass of the root class defined by the $Thing$ abstract class, both modelled as sets. The $subclass$ $relationship$ is defined as a set inclusion relationship (encoding a subsumption relationship) between the corresponding sets to the subclass and the mother class, and the equivalence relationship is defined in Event-B using the set equality relationship between the corresponding sets to the equivalent classes. The union combination of two classes is modeled in Event-B as the set union of the two sets corresponding to the two classes. To get all formalisation rules defined for the shallow modelling process, the reader may refer to this reference \cite{mediMohandOussaidA17}.

By applying some of this formalisation rules to the diplomas ontology described in section \ref{methods}, we get the following Event-B context (figure \ref{fig:modele1}).

\begin{figure}[!h]
\begin{center}
\scriptsize{
\begin{tabular}{|p{12cm}|}
\hline
\begin{description}
\CONTEXT{$Ontology$}
\SETS
	\begin{description}
	\ItemX{Thing }
	\end{description}
\CONSTANTS
	\begin{description}
	\ItemX{Phd~Master~Engineer~Diplom~Bachelor~Diplomas\_For\_Phd}
	\end{description}
\AXIOMS
	\begin{description}
	\nItemX{axm1}{Diplom \subseteq Thing}
	\nItemX{axm2}{Bachelor \subseteq Diplom}
	\nItemX{axm3}{Master \subseteq Diplom}
	\nItemX{axm4}{Engineer \subseteq Diplom}
	\nItemX{axm5}{Engineer = Master}
	\nItemX{axm6}{Phd \subseteq Diplom}
	\nItemX{axm7}{Diplomas\_For\_Phd = (Engineer \cup Master)}
	\end{description}
\END
\end{description}
\\
\hline
\end{tabular}}
\caption{Event-B context for diplomas : shallow modelling}
\label{fig:modele1}
\end{center}
\end{figure}

%%%%%%%%%%%%%%%%%%%%%%%%%%%%%%%%%%%%%%%
%%%%%%%%%%%%%%%%%%%%%%%%%%%%%%%%%%%%%%%
\subsection{Deep modelling: Ontologies as instances of ontology models}

The approach that uses deep modelling consists in modelling the ontology concepts together with the concepts of the modelling language that were used to define the ontology concepts \cite{mediHacidA16,Isola2016}. Here, ontologies are defined as instances of ontology models. Two steps are required. First, an ontology model is formalised and then ontologies are defined as specific models corresponding to the defined ontology model. In our approach, we consider that both ontology modelling concepts and ontologies are explicitly modelled.

We have used the Event-B method to formalise these concepts. More precisely, as we consider ontologies as theories, we have used Event-B contexts to formalise such concepts. Classes, properties, instances and values are defined by the $CLASS$, $PROPERTY$, $INSTANCE$ and $VALUE$ carrier sets. These sets are abstractly defined, they are  populated when defining specific ontologies.

Several relationships available in ontology modelling languages have been formalised. We have  modelled $subclass$ as a relation between classes. A set $IS\_A$ gathers the possible $subclass$ relations between classes.  A second part of this definition describes the constraints associated to inheritance i.e. inclusion of sets of instances. Indeed, in $axm2$ of figure \ref{fig:modele2}, it is explicitly stated that the set of instances of a class $x$ such that $x$ $Is\_a$ y is included in the set of instances of class $y$. 

To model the equivalence relationship, we proceed in the same manner as for the $Is\_a$ relationship. First, the equivalence is a relation between classes. Second, the axiom $axm3$ states that the defined relation is reflexive, symmetric and transitive. 

The $UnionOf$ operator is defined as a relation between sets of classes. The defined logical property states that if an instance belongs to a class $x$ or an instance belongs to a class $y$ then it belongs to the class $z$ belonging to the $UnionOf$ relation (axiom $axm4$ of figure \ref{fig:modele2}). 

To obtain the definitions of  all the formalisation rules defined for the deep modelling process, the reader may refer to these references \cite{mediHacidA16,Isola2016}.

\begin{figure}[!h]
\begin{center}
\scriptsize{
\begin{tabular}{|p{14cm}|}
\hline
\begin{description}
\CONTEXT{$Ontology\_Model$}
\SETS
	\begin{description}
	\ItemX{CLASS~PROPERTY~INSTANCE~VALUES~\ldots }
	\end{description}
\CONSTANTS
	\begin{description}
	\ItemX{HAS\_INSTANCES~\ldots~IS\_A~\ldots~EQUIVALENCE~\ldots~UNION\_OF~\ldots}
	\end{description}
\AXIOMS
	\begin{description}
	\nItemX{axm1}{HAS\_INSTANCES = CLASS \rel  INSTANCE}
	\nItemX{axm2}{IS\_A =\{IsA | IsA \in CLASS \rel CLASS \land (\forall x, y\qdot(x \in CLASS \land y \in CLASS \land x \mapsto y \in IsA \Leftrightarrow union(\{r \qdot r \in HAS\_INSTANCES| ran(\{x\} \triangleleft r)\})   \subseteq union(\{r \qdot  r  \in  HAS\_INSTANCES| ran(\{y\} \triangleleft r)\}) ))	)\}}
	\nItemX{axm3}{EQUIVALENCE = \{ EQo |  EQo \in  CLASS \rel  CLASS  \land (\forall x \qdot  (x \in  CLASS \limp  x\mapsto x \in  EQo)) \land (\forall  x, y \qdot  (x \in  CLASS \land  y \in CLASS  \land  x \mapsto y \in  EQo \limp  y \mapsto x \in  EQo)) \land (\forall  x, y, z \qdot  (x \in  CLASS \land  y \in  CLASS \land  z \in  CLASS \land  x \mapsto y \in  EQo \land  y  \mapsto z \in  EQo \limp  x \mapsto z \in  EQo))\}}
	\nItemX{axm4}{UNION\_OF = \{unionOf| (unionOf \in (\pow(CLASS) \cprod \pow(CLASS) \rel CLASS))  \land (\forall x, y, z\qdot(x \in \pow(CLASS) \land y \in \pow(CLASS) \land z \in CLASS \land x\mapsto y \mapsto z \in unionOf \limp \forall instance\qdot (instance \in INSTANCE \limp \exists hasInstance\qdot (hasInstance \in HAS\_INSTANCES \limp (\forall n, m\qdot (n \in x \land m \in y \land (n\mapsto instance \in hasInstance \lor m\mapsto instance \in hasInstance)) \limp z \mapsto instance  \in  hasInstance)))) ) \}}
	\nItemX{axm\_i}{...}
	\end{description}
\END
\end{description}
\\
\hline
\end{tabular}}
\caption{Event-B generic context for ontology : deep modelling}
\label{fig:modele2}
\end{center}
\end{figure}

In figure \ref{fig:modele3}, we give an extract of the ontology of diplomas we have formalised as instances of the generic concepts previously introduced. The defined ontology illustrates the $subClassOf$, $Equivalence$ and $UnionOf$ relationships.

\begin{figure}[!h]
\begin{center}
\scriptsize{
\begin{tabular}{|p{14cm}|}
\hline
\begin{description}
\CONTEXT{$Diplomas\_Ontology$}
\EXTENDS{$Ontology\_Model$}
\CONSTANTS
	\begin{description}
	\ItemX{Diplom~Bachelor~Master~Engineer~Phd~Diplomas\_For\_Phd}
	\ItemX{isA~eQ~unionOf}
	\end{description}
\AXIOMS
	\begin{description}
	\nItemX{axm1}{partition(CLASS,\{Diplom\},\{Bachelor\},\{Master\},\{Engineer\},\{Phd\},\{Diplomas\_For\_Phd\})}
	\nItemX{axm2}{isA = \{
		Master \mapsto Diploms, 
		Bachelor \mapsto Diploms, 
		Engineer \mapsto Diploms, Phd \mapsto Diploms
			\} }
	\nItemX{axm3}{eQ = \{
		Bachelor \mapsto Bachelor, Master \mapsto Master, 
		Engineer \mapsto Engineer, Phd \mapsto Phd, 
		Master \mapsto Engineer, Engineer \mapsto Master
			\} }
	\nItemX{axm4}{unionOf = \{
  		\{Master\} \mapsto \{Engineer\}  \mapsto  Diplomas\_For\_Phd
  		\}}
	\nItemX{axm\_i}{...}
	\end{description}
\THEOREMS
	\begin{description}
	\nItemX{thm1}{isA \in IS\_A }
	\nItemX{thm2}{eQ  \in EQUIVALENCE}
	\nItemX{thm3}{unionOf  \in UNION\_OF}
	\end{description}
\END
\end{description}
\\
\hline
\end{tabular}}
\caption{Event-B context for diplomas : deep modelling}
\label{fig:modele3}
\end{center}
\end{figure}

%%%%%%%%%%%%%%%%%%
%%%%%%%%%%%%%%%%%%
\section{The OntoEventB plug In}

\begin{figure}[!h]
\begin{center}
\includegraphics[scale=0.45]{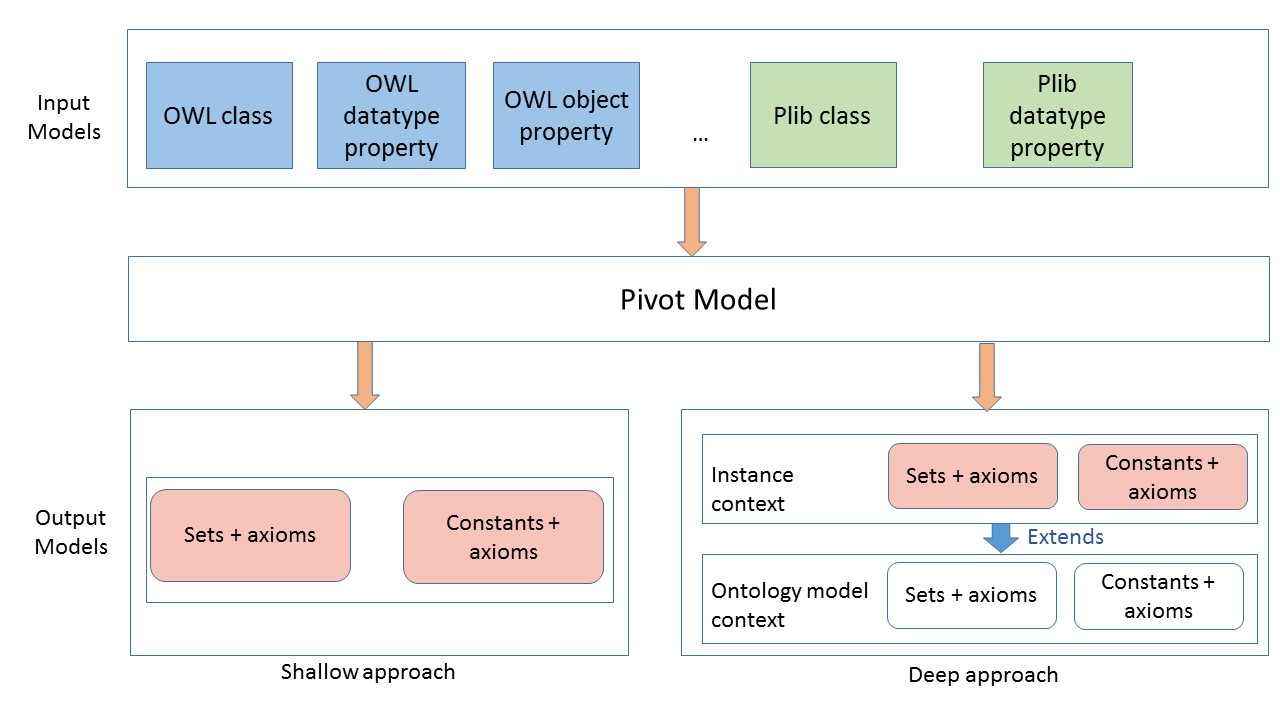}
\end{center}
\caption{The OntoEventB internal architecture.}
\label{OntoEventB}
\end{figure}

The OntoEventB plug-in \cite{afadlMohandOussaidA17}  has been developed to automatically support the translation of ontologies models, described using ontology description languages such as OWL \cite{owl2-overview} or PLIB \cite{PLIB}, into Event-B Contexts \cite{Abrial10}. It takes as input an ontology description file and generates, according to the selected approach (shallow or deep), the corresponding Event-B contexts. The OntoEventB plug-in is developed according to an architecture composed of three components: Input, Pivot and Output Models (Figure \ref{OntoEventB}).

\paragraph{The Input Models component.}This component is devoted to the processing of the input models described using different ontology description languages such as OWL, PLIB ... It browses the input models files in order to extract ontological concepts descriptions (e.g. OWL classes, OWL data type properties and OWL object properties in the case of OWL models) and to send them to the Pivot
Model component.

\paragraph{The Pivot Model component.}  This component is an intermediate operational model, which summarizes the common relevant concepts used by ontology description languages (classes, properties and data types). It defines generic concepts that integrate all specific concepts that can be received from the Input Model component. The Pivot Model can be extended to integrate other generic concepts that can be identified if a new language is added as input model in the Input Models component.

When different ontological concepts are produced from Input Model components (e.g.OWL classes, OWL data type properties and OWL object properties in the case of OWL models), the Pivot Model component translates them into its generic concepts (classes, properties and data types). After this first translation step, the obtained generic concepts are ready to be treated by the next process handled by the Output Model component.

\paragraph{The Output Model component.} This component has as input the generic concepts computed by the Pivot Model component and translates them into Event-B Context elements (sets, constants and axioms). This process uses transformation rules that formalise each ontological concept by an Event-B definition following the two approaches proposed and described in section \ref{theories} (Shallow and Deep modelling approaches). The user of the OntoEventB plug-in can choose one of them.

The use of this architecture allows us to extend the OntoEventB plug-in by taking into account new
input ontology description languages without redefining the Event-B formalisation rules between Pivot
Model component and Output Model component. Indeed, as soon as the new
concepts defined by these new languages are translated into generic
concepts of the Pivot model, they are be directly formalised in
the Event-B Context elements without redefining new transformation rules.

\paragraph{Installing and Using OntoEventB plug-in.}
\begin{figure}[!h]
\begin{center}
\includegraphics[scale=0.65]{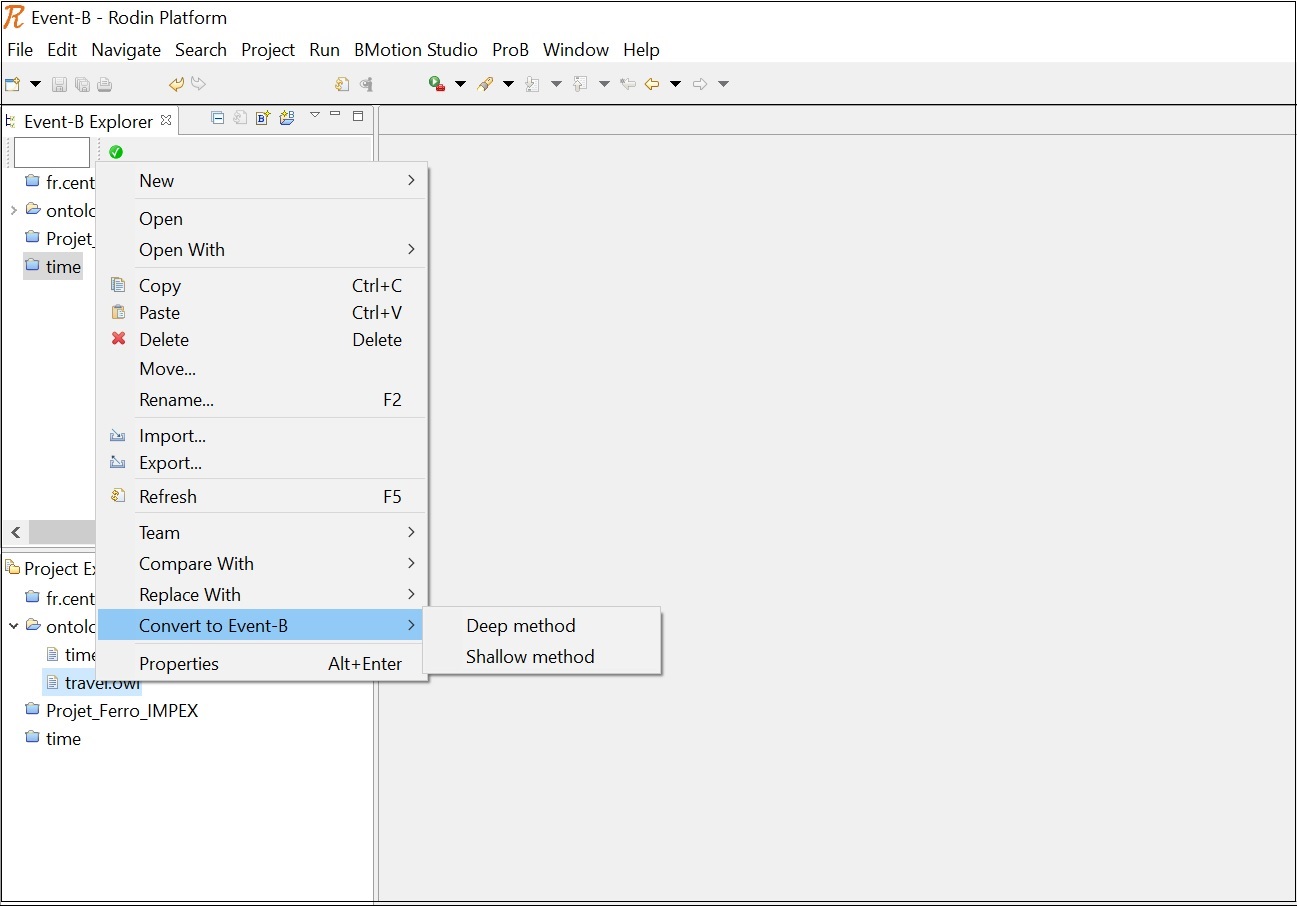}
\end{center}
\caption{The OntoEventB submenu.}
\label{OntoEventBSM}
\end{figure}

The OntoEventB tool is developed as an Eclipse plug-in to integrate it into a Rodin platform \cite{rodin}, an IDE (Integrated Development Environment) supporting Event-B developments. To use OntoEventB plug-in in your Rodin platform instance, you must install the plug-in by using the Install New Software menu item\footnote{OntoEventB update site : \url{http://wdi.supelec.fr/OntoEventB-update-site/}} for downloading and installing the plug-in automatically.

After installing the OntoEventB plug-in in a Rodin platform instance, the convert to Event-B sub-menu becomes available by right clicking on an owl file (with an\texttt{.owl} extension) in the project explorer as shown in Figure \ref{OntoEventBSM}. It proposes to se up the two modelling techniques: deep and shallow  corresponding to the two proposed approaches we introduced section \ref{theories}.

%%%%%%%%%%%%%%%%%%
%%%%%%%%%%%%%%%%%%  
\section{Conclusion}
This paper reports on some of the results of the French ANR IMPEX research project. We have discussed the interest of making explicit domain knowledge in design models in order to strengthen them. We also proposed a straightforward approach formalizing ontologies as theories encoded within Event-B contexts. This approach led to the development of Plug-In that produces automatically Event-B contexts from ontologies expressed in different ontology models.

Moreover, the previous work achieved in this project showed the interests of the approach to strengthen models in different areas. We have applied the developed approach to case studies issued from avionic systems, medical devices and electronic voting systems.

This work is still an on-going work. We are currently investigating the possibility to formalise ontologies of behaviours (e.g. ontologies of services) and their use to annotate behavioural components of design models (e.g. events of an Event-B model). First results are already available on plastic interfaces \cite{tase15}. 

Other investigations consider design system models re-factoring with the objective  of handling explicitly domain knowledge in order to support the verification of new properties mined from domain models.

\bibliographystyle{eptcs}
\bibliography{idir}
\end{document}